# Pressure-induced metallization and superconducting phase in ReS$_2$

Dawei Zhou[1,2,*], Yonghui Zhou[3,*], Chunying Pu[2], Xuliang Chen[3], Pengchao Lu[1], Xuefei Wang[3], Chao An[3], Ying Zhou[3], Feng Miao[1], Ching-Hwa Ho[4], Jian Sun[1,5,†], Zhaorong Yang[3,5], Dingyu Xing[1,5]

[1] *School of Physics and National Laboratory of Solid State Microstructures, Nanjing University, Nanjing 210093, China*

[2] *College of Physics and Electronic Engineering, Nanyang Normal University, Nanyang 473061, China*

[3] *High Magnetic Field Laboratory, Chinese Academy of Sciences, Hefei 230031, China*

[4] *Graduate Institute of Applied Science and Technology, National Taiwan University of Science and Technology, Taipei 106, Taiwan*

[5] *Collaborative Innovation Center of Advanced Microstructures, Nanjing University, Nanjing 210093, China*

**Abstract**：Among the family of TMDs, ReS$_2$ takes a special position, which crystalizes in a unique distorted low-symmetry structure at ambient conditions. The interlayer interaction in ReS$_2$ is rather weak, thus its bulk properties are similar to that of monolayer. However, how does compression change its structure and electronic properties is unknown so far. Here using *ab initio* crystal structure searching techniques, we explore the high-pressure phase transitions of ReS$_2$ extensively and predict two new high-pressure phases. The ambient pressure phase transforms to a "distorted-1T" structure at very low pressure and then to a tetragonal $I4_1/amd$ structure at around 90 GPa. The "distorted-1T" structure undergoes a semiconductor-metal transition (SMT) at around 70 GPa with a band overlap mechanism. Electron-phonon calculations suggest that the $I4_1/amd$ structure is superconducting and has a critical superconducting temperature of about 2 K at 100 GPa. We further perform high-pressure electrical resistance measurements up to 102 GPa. Our experiments confirm the SMT and the superconducting phase transition of ReS$_2$ under high pressure. These experimental results are in good agreement with our theoretical predictions.

*D.W.Z and Y.H.Z contributed equally to this work. †Correspondence should be addressed to: J.S. (jiansun@nju.edu.cn)

**INTRODUCTION**

The transition metal dichalcogenides (TMDs) $MX_2$ (M, transition metal; X, chalcogen, S, Se, Te) have become a topic of current interest in recent years due to their unique structural, electronic, optical, and tribological properties [1-12]. Up to now, there are approximately more than 60 TMDs reported, and most of them have a layered structure resembling graphite, although the layers are actually different. In the layers of TMDs, a transition metal atom layer is sandwiched by two chalcogen atom layers, forming a X-M-X sandwiched structure. The X-M-X trilayers then stack together via weak van der Waals interaction, resulting in the different phases in TMDs such as 1T, 1T', $2H_c$, $2H_a$ and 3R [12-17]. Due to their unique layered structure, TMDs exhibit some exceptional functional properties under pressure such as superconductivity [3, 18-20] and charge density wave (CDW) [6, 21-23]. Moreover, as a result of the weak interlayer interactions, similar to graphite, their bulk materials can be cleaved easily into monolayer, forming a new class of two dimensional (2D) materials with many novel properties [24, 25]. For example, these cleaved monolayers can be reassembled layer by layer with a specified sequence to form programmed heterostructures and devices [26]. Using so called "vertical" techniques, such van der Waals heterostructures with unique properties have been fabricated recently [27, 28].

More interestingly, compared with the bulk form, monolayer and few-layer TMDs generally exhibit distinct electronic structure and optical properties. For example, the band gaps of some semiconducting TMDs will change from indirect to direct in single layer [25, 27]. Among the family of TMDs, $ReS_2$ takes a special position and has received incremental attention recently [30, 31]. Unlike most of the TMDs, the ambient $ReS_2$ crystallizes in a distorted CdCl structure with low triclinic symmetry [32, 33] (space group: *P*-1, although this structure was usually named as 1T' in the previous literatures, to describe the structural evolution of $ReS_2$ clearly, as discussed later, we name this ambient structure "distorted-3R" in this paper). The low symmetry structure and the distortion of Re and S atoms were related to the Peierls mechanism [31, 34], which prevents ordered stacking and minimizes the interlayer overlap of wave functions. Thus the interlayer interaction in $ReS_2$ is rather weak, and the bulk properties are similar to that of monolayer [31, 35]. For instance, the direct-to-indirect band gap transition as reported in $MoS_2$ is not observed in this material when going from the monolayer to bulk [31].

Since the electronic and optical properties do not change so much in $ReS_2$ with increasing layers, applying stress and pressure becomes an important approach to modify its properties. Generally speaking, pressure can directly modify the lattice parameters, changing effectively its electronic structures and optical properties. It can also symmetrize the structure and vanish the Peierls distortion. Furthermore, as a member of TMDs, metallization, CDW and possible superconductivity under pressure are also important issues. Among above-mentioned issues, it is the key point to figure out the structural evolution of $ReS_2$ under pressure. In the previous work, Hou *et al.* [36] have explored the high-pressure structures of $ReS_2$ up to 51 GPa using synchrotron x-ray diffraction (XRD), they found an indication of the occurrence of a new high-pressure phase at 11 GPa. Kao *et al.* [37] also found a transition in $ReSe_2$ at the pressure of 10.5 GPa. However, both of them did not resolve the crystal structure of the high-pressure phases. In this paper, using efficient crystal structure search methods combining with *ab initio* calculations, we predict the evolution of the structural and electronic properties of $ReS_2$ under pressure. Two new high-pressure phases are identified, and the structural characteristics and their stability, electronic structure, metallization of $ReS_2$ under pressure are investigated systematically. We also perform high-pressure electrical transport measurements up to 102 GPa, which confirm our predictions on the metallization and superconductivity very well. Unlike the most TMDs which undergo a phase transition from $H_c$ to $H_a$ under pressure, we find that $ReS_2$ has a different phase transition sequence under pressure. Beside the known structure types of 3R, 1T, 4H, 1T, several new crystal structures of TMDs family have been proposed in our work. Our results reveal that pressure plays an important role in modifying and symmetrizing the crystal structure in the distorted TMDs system indeed.

**RESULTS**

Structure predictions were carried out with simulation cells containing 1−8

formula units (f.u.) of $ReS_2$ under pressure from 0 GPa to 200 GPa. The experimentally observed ambient structure (distorted-3R phase) was successfully reproduced at 0 GPa, validating our methodology. To get accurate results, van der Waals (vdW) interaction is taken into consideration. The optB88 functional together with the vdW-DF corrections [38] as implemented in VASP was concerned during calculating the enthalpy-pressure, due to their good performance in the layered TMDs [39]. The enthalpy of the typical known structures in TMDs and our predicted ones is shown in Fig. 1 (a). Compared with the experimental lattice parameters of the ambient distorted-3R structure, the optB88+vdW-DF functionals gives an error within only 3%, indicating that this computational setting can give good descriptions of vdW forces in $ReS_2$ (see supplementary Table S1). We also calculated the enthalpy of the newly found structure relative to distorted-3R phase using LDA and PBE exchange correlation functionals without vdW corrections, which is shown in the inset of Fig. 1 (a). At about 0.1 GPa, a new energetically favorable structure with the same triclinic symmetry (space group: *P*-1, we name it "distorted-1T") is uncovered, the supercell of this structure is shown in Fig. 1(b). Although the phase transition pressure changes from 0.4 GPa for LDA to 3 GPa for PBE, we can conclude that the newly found structure becomes more energetically favorable than the ambient distorted-3R structure above pretty low pressure and is stable in a wide pressure range to about 90 GPa. After that, a tetragonal *I*4$_1$/*amd* structure is found to be more preferable. The detailed structural parameters of these two predicted phases are summarized in the supplementary (see supplementary Table S2). The calculated phonon dispersions of these two newly found phases are shown in supplementary Fig.S1. It is shown that both phases are dynamically stable proved by the absence of any imaginary phonon frequency in the whole Brillouin zone. The volume changing with pressure is also calculated, as shown in supplementary Fig.S2. With increasing pressure, the volume changes of distorted-3R and distorted-1T are slightly different, while a discontinuous volume change occurs in the transition from distorted-1T to *I*4$_1$/*amd*, which exhibits a first order phase transition.

The newly found distorted-1T has the same space group with the experimentally observed ambient distorted-3R, both belong to *P*-1, however, their packing is different. During optimizing these two structures under higher pressure, we find that the previously known ambient pressure *P*-1 structure tends to symmetrize into 3R structure (space group: R-3m) while the new *P*-1 structure found in this work tends to symmetrize into 1T structure (space group: *P*-3*m*1), respectively, (see Fig. 1 (b) and Fig. 1 (c)). That is the reason why we think it is more proper to name the known ambient *P*-1 structure as "distorted-3R" although it is usually called "distorted-1T" or 1T' in the literature before. Correspondingly, the newly found *P*-1 structure in this work is named as "distorted-1T" for the same reason. Interestingly, the 3R and 1T structures were reported to be found in other TMDs such as $MoS_2$. For $MoS_2$, 1T phase is metastable and metallic [14, 15], while 3R phase is semiconducting and it keeps stable under high pressure and high temperature [16, 17]. The main difference between the 3R and 1T is the stacking sequence of the S-Re-S trilayer. Different from 3R phase, all Re atoms in 1T phase share the same (x, y) coordinate. The 3R structure can transform to the 1T structure with a relative sliding between sandwiched layers. From the analysis above, one can find that the layer-sliding might also be responsible for the transition from the distorted-3R to distorted-1T phase. In fact, similar layer-sliding has been observed in other TMDs under pressure such as $MoS_2$, which leads to the transition from $2H_c$ to $2H_a$ [13], while in $WTe_2$, it is from Td to 1T' and finally 2H [39].

Previously, Hou *et al.* [36] have explored $ReS_2$ up to 51 GPa using X-ray powder diffractions (XRD) with a wavelength of 0.4959 Å, and reported a high-pressure phase at around 11 GPa, which is stable at least up to 51 GPa. However, they did not resolve the crystal structure of this high-pressure phase because of the peak broadening and overlapping. According to the experiment, the phase transition is indicated by the remarkable change of the (220) peak, the relative intensity of which considerably increased when the phase transition occurred. In Fig.2, we show the simulated XRD curves of distorted-3R and distorted-1T structures together with the experimental results at 11GPa and 50 GPa, respectively. As one can see, the XRD curves of our predicted distorted-1T structure agree well with the experimental results.

Particularly, the remarkable change of the special peak during the phase transition matches the experimental results nicely. We find that if using a longer x-ray wavelength, for instance λ=1.54056Å, to simulate the XRD curves, the XRD patterns of distorted-3R and distorted-1T structures will show more difference and easy to be distinguished, especially for the peaks between 35 and 40 degree (see supplementary Fig.S3(a) for more details). Furthermore, as shown in Fig. S3, the lattice parameters of these two *P*-1 structures show different trend, indicating that they have quite different anisotropy under pressure. However, the experimental results indicate that the new phase emerges at around 11 GPa, which is higher than our theoretical transition pressure. The difference might be due to a large activation barrier for the phase transition.

## DISSCUSSIONS

The distortion of S and Re atoms in distorted-3R phase was believed to be related to the Peierls mechanism [34]. The distortion reduces the band energy from electrons, while increases the repulsion energy between the atoms. However, if the decrease of the band energy is sufficient enough to overcome the cost in increased repulsion, the distorted low-symmetry structure can exist stably. Generally speaking, pressure tends to destroy the Peierls distortion, and leads to high-symmetry structure. To reveal the origin of the atoms distortion in the distorted-1T structure, we show the total density of states (TDOS) of the distorted and undistorted 3R and 1T phases at 15 GPa in Fig.3(a)-(b). It can be seen that the distortion introduces a band gap in both 3R and 1T structures, as this greatly decreases the band energy. Particularly, we noticed that, through the layer-sliding, the first peak of total DOS below the Fermi level in distorted-1T phase is lower than that of distorted-3R phase, thus the distorted-1T phase decreases more energy than the distorted-3R phase. Therefore, we can see that the Peierls mechanism still plays an important role in the newly found distorted-1T phase. The partial density of states (PDOS) of distorted-1T phase at 15 GPa are also show in Fig. 3 (b), both the valence band maximum (VBM) and the conduction band minimum (CBM) are mainly derived from hybridized Re-d and S-p states, while other states such as S-s, S-d and Re-p have negligible contributions to the VBM and CBM.

We further investigated the band structure of distorted-1T phase at different pressures. The spin-orbit interactions are taken into account during the band structure calculations, see Fig. 3 (d) in details. At 15 GPa, both VBM and CBM are located at the Q point, so it is a direct bandgap semiconductor at 15 GPa, and the bandgap is about 0.66 eV. As pressure increases, the lowest line of the conduction band moves down, the eigenvalues at B point and the one between Z and G point reduce quite a bit. Both of them become lower than the Q point state, leading to a transition from direct gap to indirect gap at about 60 GPa. Upon further compression, the distorted-1T structure undergoes a semiconductor-metal phase transition at about 70 GPa, which results from the closure of the band gap. The detailed changes of the band gap with pressure are shown in supplementary Fig.S4.

Under higher pressure, a new 3D highly compact phase with tetragonal $I4_1/amd$ symmetry is predicted to be more energetically stable, taking the place of sandwich-layered structures, see Fig. 4 (a) in detail. From the band structure and total density of states in Fig. 4 (c), we can see that tetragonal $I4_1/amd$ phase is strongly metallic. Moreover, the metallic character of this structure is mainly contributed by Re-d states with fewer contributions from the S-p states. We further calculated the electron-phonon coupling (EPC), phonon spectra, partial phonon DOS, Eliashberg EPC spectral function $a^2F(\omega)$ and the electron-phonon integral $\lambda(\omega)$ at 100 GPa of this phase with density functional perturbation theory [40], which are shown in Fig.4(d). We found the EPC constant $\lambda$ to be about 0.41. The estimated critical temperature superconductivity ($T_c$), using the Allen-Dynes equation [41] with $\mu^* = 0.1$ is about 2 K, which agrees with our experimental results reasonably well as discussed later. From the calculation of the phonon linewidths, it can be seen that the intermediate- and high-frequency S vibrations make a significant contribution to the overall EPC constant. The partial DOS, Eliashberg EPC spectral function $a^2F(\omega)$ and the electron-phonon integral $\lambda(\omega)$ also confirm this. It seems that S atoms may be vitally important to the superconductivity of $ReS_2$.

To experimentally confirm the predicted semiconductor-semimetal phase transition and the superconducting phase, we investigate the evolution of resistance as a function of temperature for $ReS_2$ single crystal under high pressure up to 102.0 GPa in Fig. 5. At 4.0 GPa, $ReS_2$ exhibits the semiconducting conductivity similar to its ambient pressure behavior. With increasing pressure up to 11.0 GPa, the overall resistance significantly decreases by two orders of magnitude. Especially at 15.1 GPa, a non-semiconducting behavior emerges above 100 K where the resistance is almost independent on temperature, which could be attributed to the pressure-induced structural transition from distorted-3R to distorted-1T as mentioned above. The pressure-induced metallic conductivity can be recognized under 32.0 GPa in the high temperature range while the resistance at low temperature still keeps the semiconducting profile. Note that the semiconducting profile preserves till 70.0 GPa, as shown in Fig. 5(b), which implies that the ground state is indeed semiconducting in nature. Only when the pressure is further enhanced to 81.0 GPa, the semiconducting behavior is almost completely suppressed. The characteristic pressure could also be traced in the pressure dependence of isothermal resistance shown in Fig. 6(a), where a small change of slope takes places around 70 GPa, in well agreement with the theoretically predicted semiconductor-metal phase transition in the distorted-1T structure.

More interestingly, in accordance with the electron phonon calculations, when the pressure reaches to 91.0 GPa, a small drop in the resistance curve appears around 2.5 K, as shown in Fig.5(c). The resistance drop becomes more and more pronounced with increasing pressure up to 102.0 GPa, the limiting pressure in our experimental set-up. To make sure that the drop of the resistance was indeed a superconducting transition, we carried out electrical resistance measurements under various external magnetic fields aligned along $c$-axis of $ReS_2$ at 102.0 GPa. As seen from Fig. 6(b), the resistance drop is gradually suppressed and moves towards low temperature curve with the increasing field up to 1.0 T. The Ginzburg-Landau (GL) fitting yields a critical field of 2.509 T, as shown in the inset of Fig. 6(b). The suppression of resistance drop by magnetic field thus confirms the presence of pressure-induced superconductivity. The failure to observe zero resistance is most likely to be caused by the non-hydrostatic compressive stress, which results in a huge pressure gradient in the sample. To achieve zero resistance in the superconducting phase with the critical temperature around 2 K, both higher pressure and lower temperature limit should be reached. Upon pressure releasing from 102.0 GPa, the superconductivity retains itself until 55.0 GPa, as shown in the inset of Supplementary Fig.S5, which is accompanied by the increasing normal state resistance.

In summary, we have explored the ground-state structures of $ReS_2$ under pressure systematically using crystal structure prediction methods combining with *ab initio* calculations. Based on structural evolution, we think it is more proper to name the ambient *P*-1 structure "distorted-3R", although it is usually called as 1T' in the literature. Upon very small compression, the ambient distorted-3R is predicted to transform to a new triclinic distorted-1T structure with the same *P*-1 symmetry. The simulated XRD patterns of this new phase agree well with the previous experimental XRD results under pressure. The layer-sliding is found to be responsible for the transition, and from the electronic structures, we find the Peierls mechanism playing an important role in decreasing the energy of the low symmetry structures in $ReS_2$. Moreover, through high-pressure transport measurements, $ReS_2$ is confirmed to undergo transitions from semiconductor to metal under pressure. Under higher pressure, a new highly compact metallic phase with tetragonal *I*$4_1$/*amd* symmetry becomes stable above 90 GPa, which has a critical superconducting temperature of about 2 K at 100 GPa. This work shows that pressure plays a vital part in changing and symmetrizing the crystal structure in the distorted TMDs system, which also leads to dramatic modifications on its electronic structures, such as band closure and superconducting transition.

**METHODS**

*Ab initio* **calculations**

Structure prediction through a global minimization of free energy surfaces has been successfully applied to predict high-pressure structures [42-46]. Here we extensively searched for ReS$_2$ ground-state structures under pressure based on the CALYPSO (Crystal structure Analysis by Particle Swarm Optimization) method [47, 48]. Furthermore, to ensure the reliability of the most stable structures, another structure searching method, *ab initio* random searching (AIRSS) [49, 50], was used to check the results at selected pressure. The geometry optimization and electronic calculations were performed in the framework of the density functional theory within the generalized gradient approximation Perdew-Burke-Ernzerhof (GGA-PBE) [51], as implemented in the Vienna *ab initio* simulation package (VASP) [52] code. The all-electron projector-augmented wave (PAW) method [53] was adopted with $2p3s$ and $4s4p5s4d$ as valence electrons for S and Re atoms, respectively. For the structure searches, a plane-wave basis kinetic energy cutoff of 400 eV and a grid of spacing $2\pi \times 0.06$ Å$^{-1}$ for Brillouin zone (BZ) sampling were found to be sufficient. While the enthalpy and electronic structure calculations were performed at a higher level of accuracy consisting of a basis energy cutoff of 800 eV and a k-point grid spacing of $2\pi \times 0.025$ Å$^{-1}$. The phonon calculations are carried out by using a supercell approach, as implemented in the phonopy code [54]. The full potential linearized augmented plane-wave (FLAPW) method implemented in WIEN2K code [55] was used to compute the band structure of the distorted-1T and $I4_1/amd$ structures at selected pressure points. The electron phonon coupling (EPC) were calculated using the QUANTUMESPRESSO code [40]. The ultrasoft Vanderbilt pseudopotentials with a Perdew-Burke-Ernzerhof (PBE) [51] exchange-correlation (XC) functional were used, and the energy cut-off of the plane wave basis was set at 45 Ry. The k-space and q-point integrations over BZ were performed on a $12 \times 12 \times 12$ grid and a $4 \times 4 \times 4$ grid, respectively.

**High-pressure experiments**

High pressures were generated with a screw-pressure-type diamond anvil cell (DAC) placed inside a homemade multifunctional measurement system (1.8-300 K, JANIS Research Company Inc.; 0-9 T, Cryomagnetics Inc.). Diamond anvils of 200-μm culets and rhenium gasket covered with a mixture of epoxy and fine cubic boron nitride (*c*BN) powder were used for both high-pressure transport-measurements. A single crystal with dimension of 70 × 15 × 5 μm$^3$ was loaded without pressure-transmitting medium. The four-probe method was applied in the *a-b* plane of single crystal. Platinum (Pt) foil with a thickness of 5 μm was used for the electrodes. Pressure was calibrated by using the ruby fluorescence scale below 70 GPa [56] and the diamond Raman scale above 70 GPa [57].

**Acknowledgements**

This research was financially supported from the MOST of China (Grant Nos. 2016YFA0300404, 2016YFA0401804 and 2015CB921202), the National Natural Science Foundation of China (Grant Nos: 51372112, 11574133, 11574323, U1332143, U1632275 and 51501093), The Henan Joint Funds of the National Natural Science Foundation of China (Grant Nos: U1304612 and U1404608), NSF Jiangsu province (No. BK20150012), the Fundamental Research Funds for the Central Universities, Special Program for Applied Research on Super Computation of the NSFC-Guangdong Joint Fund (the 2nd phase) and Open Fund of Key Laboratory for Intelligent Nano Materials and Devices of the Ministry of Education (INMD-2016M01). The Postdoctoral Science Foundation of China (Grant No. 2015M581767). Young Core Instructor Foundation of Henan Province (No. 2015GGJS-122), Science Technology Innovation Talents in Universities of Henan Province (No.16HASTIT047). Part of the calculations was performed on the supercomputer in the HPCC of Nanjing University and "Tianhe-2" at NSCC-Guangzhou.

**Competing financial interests:**
The authors declare no competing financial interests.

**Additional information**

Supplementary Information accompanies this paper at:
http://www.nature.com/naturecommunications

## Contributions

J.S. conceived the project; D.W.Z., P.C.L. and C.Y.P. performed calculations with the supervision of J.S.; Y.H.Z., X.L.C., X.F.W., C.A. and Y.Z. performed the high pressure experiments with the supervision of Z.R.Y.; D.W.Z, Y.H.Z., P.C.L., J.S. and Z.R.Y. analyzed the data; F.M. and C.H.H. provided the $ReS_2$ sample; D.W.Z., Y.H.Z., J.S., Z.R.Y., D.Y.X. wrote the manuscript. All authors discussed the results and commented on the manuscript.

## REFERENCE

[1] Chi, Z. H. *et al.* Pressure-Induced Metallization of Molybdenum Disulfide. *Phys. Rev. Lett.* **113**, 036802 (2014).

[2] Kang D. F. *et al.* Superconductivity emerging from a suppressed large magnetoresistant state in tungsten ditelluride. *Nat. Commun.* **6**, 7804 (2015)

[3] Pan, X.C. *et al.* Pressure-driven dome-shaped superconductivity and electronic structural evolution in tungsten ditelluride. *Nat. Commun.* **6**, 7805 (2015).

[4] Zhao, Z. *et al.* Pressure induced metallization with absence of structural transition in layered molybdenum diselenide. *Nat. Commun.* **6**, 7312 (2015).

[5] Chhowalla, M. *et al.* The chemistry of two-dimensional layered transition metal dichalcogenide nanosheets. *Nat. Chem.* **5**, 263 (2013).

[6] Barnett, R. L. *et al.* Coexistence of Gapless Excitations and Commensurate Charge-Density Wave in the 2H Transition Metal Dichalcogenides. *Phys. Rev. Lett.* **96**, 026406 (2006).

[7] Thoutam, L. R. *et al.* Temperature-Dependent Three-Dimensional Anisotropy of the Magnetoresistance in $WTe_2$. *Phys. Rev. Lett.* **115**, 046602 (2015).

[8] Yuan, N. F. Q., Mak, K. F. & Law, K. T. Possible Topological Superconducting Phases of $MoS_2$. *Phys. Rev. Lett.* **113**, 097001 (2014).

[9] Winer, W. O. Molybdenum disulfide as a lubricant: A review of the fundamental knowledge. *Wear* **10**, 422-452 (1967).

[10] Monney, G., Monney, C., Hildebrand, B., Aebi, P. & Beck, H. Impact of Electron-Hole Correlations on the 1T−TiSe$_2$ Electronic Structure. *Phys. Rev. Lett.* **114**, 086402 (2015).

[11] Miwa, J. A. *et al.* Electronic Structure of Epitaxial Single-Layer MoS$_2$. *Phys. Rev. Lett.* **114**, 046802 (2015).

[12] Katzke, H., Tolédano, P. & Depmeier, W. Phase transitions between polytypes and intralayer superstructures in transition metal dichalcogenides. *Phys. Rev. B* **69**, 134111 (2004).

[13] Hromadová, L., Martoňák, R. & Tosatti, E. Structure change, layer sliding, and metallization in high-pressure MoS$_2$, *Phys. Rev. B* **87**, 144105 (2013).

[14] Wypych, F. & Schollhorn, R. 1T-MoS$_2$, A New Metallic Modification of Molybdenum Disulfide, *J. Chem. Soc., Chem. Commun.* **19**, 1386-1388 (1992).

[15] Enyashin, A. N. *et al.* New Route for Stabilization of 1T-WS$_2$ and MoS$_2$ Phases. *J. Phys. Chem. C* **115**, 24586 (2011).

[16] Merrill, L. Behavior of the AB$_2$-Type Compounds at High Pressures and High Temperatures. *J. Phys. Chem. Ref. Data* **11**, 1005 (1982).

[17] Silverma, M. S. Ultrahigh pressure-high temperature synthesis of rhombohedral dichalcogenides of molybdenum and tungsten. *Inorg. Chem.* **6**, 1063 (1967).

[18] Weber, F. *et al.* Electron-Phonon Coupling and the Soft Phonon Mode in TiSe$_2$. *Phys. Rev. Lett.* **107**, 266401 (2011).

[19] Guillamón, I. *et al.* Superconducting Density of States and Vortex Cores of 2H-NbS$_2$. *Phys. Rev. Lett.* **101**, 166407 (2008)

[20] Suderow, H., Tissen, V. G., Brison, J. P., Martínez, J. L., & Vieira, S. Pressure Induced Effects on the Fermi Surface of Superconducting 2H-NbSe$_2$, *Phys. Rev. Lett.* **95**, 117006 (2005).

[21] Shen, D. W. *et al.* Novel Mechanism of a Charge Density Wave in a Transition Metal Dichalcogenide. *Phys. Rev. Lett.* **99**, 216404 (2007).

[22] Calandra, M. & Mauri, F. Charge-Density Wave and Superconducting Dome in TiSe$_2$ from Electron-Phonon Interaction. *Phys. Rev. Lett.* **106**, 196406 (2011).

[23] Chatterjee, U. *et al.* Emergence of coherence in the charge-density wave state of

2H-NbSe$_2$. *Nat. Commun.* **6**, 6313 (2015).

[24] Mak, K.F., Lee, C., Hone, J., Shan, J. & Heinz, T. F., Atomically Thin MoS2: A New Direct-Gap Semiconductor. *Phys. Rev. Lett.* **105**, 136805 (2010).

[25] Radisavljevic, B., Radenovic, A., Brivio, J., Giacometti, V. & Kis, A. Single-layer MoS$_2$ transistors. *Nat. Nanotechnol.* **6**, 147 (2011).

[26] Geim, A. K. & Grigorieva, I. V., Van der Waals heterostructures. *Nature* **499**, 419 (2013).

[27] Ponomarenko, L. A. *et al.* Tunable metal-insulator transition in double-layer graphene heterostructures. *Nat. Phys.* **7**, 958 (2011).

[28] Georgiou, T. *et al.* Vertical field-effect transistor based on graphene-WS$_2$ heterostructures for flexible and transparent electronics. *Nat. Nanotechnol.* **8**, 100 (2013).

[29] Splendiani, A. *et al.* Emerging Photoluminescence in Monolayer MoS$_2$. *Nano Lett.* **10**, 1271 (2010).

[30] Liu, E. F. *et al.* Integrated digital inverters based on two-dimensional anisotropic ReS$_2$ field-effect transistors. *Nat. Commun.* **6**, 6991 (2015).

[31] Tongay, S. *et al.* Monolayer behaviour in bulk ReS$_2$ due to electronic and vibrational decoupling. *Nat. Commun.* **5**, 3252 (2014).

[32] Murray, H. H., Kelty S. P., Chianelli, R. R. & Day, C. S. Structure of rhenium disulfide. *Inorg. Chem.* **33**, 4418 (1994).

[33] Lamfers, H. J., Meetsma, A., Wiegers, G. A. & de Boer, J. L. The crystal structure of some rhenium and technetium dichalcogenides. *J. Alloys Compd.* **241**, 34 (1996).

[34] Kertesz, M., Hoffmann, R. Octahedral vs. trigonal-prismatic coordination and clustering in transition-metal dichalcogenides. *J. Am. Chem. Soc.* **106**, 3453 (1984).

[35] Feng, Y., Zhou, W., Wang, Y., Zhou, J., Liu, E., Fu, Y., Ni, Z., Wu, X., Yuan, H., Miao, F., Wang, B., Wan, X. & Xing, D. Raman vibrational spectra of bulk to monolayer ReS$_2$ with lower symmetry. *Phys. Rev. B* **92**, 054110 (2015).

[36] Hou, D. B. *et al.* High pressure X-ray diffraction study of ReS$_2$. *J. Phys. Chem. Solids* **71**, 1571 (2010).

[37] Kao, Y. C. *et al.* Anomalous structural phase transition properties in ReSe$_2$ and Au-doped ReSe. *J. Phys. Chem.* **137**, 024509 (2012).

[38] Thonhauser, T. *et al.* Van der Waals density functional: Self-consistent potential and the nature of the van der Waals bond. *Phys. Rev. B* **76**, 125112 (2007)

[39] Lu, P. C. *et al*, Pressure-induced Td to 1T'structural phase transition in WTe$_2$ *arXiv*:1512.00604.

[40] Giannozzi, P *et al.*, *J. Phys.: Condens. Matter* **21**, 395502 (2009), http://www.quantum-espresso.org

[41] Allen, P. & Dynes, R. Transition temperature of strong-coupled superconductors reanalyzed. *Phys. Rev. B* **12**, 905 (1975).

[42] Li, Y.W., Hao, J., Liu, H.Y., Lu, S.Y. & Tse, J., High-Energy Density and Superhard Nitrogen-Rich B-N Compounds. *Phys. Rev. Lett.* **115**, 105502 (2015).

[43] Li, Y.W. *et al.* Metallic Icosahedron Phase of Sodium at Terapascal Pressures. *Phys. Rev. Lett.* **114**, 125501(2015).

[44] Li, Q., Zhou, D., Zheng, W.T., Ma, Y.M. & Chen, C.F. Global Structural Optimization of Tungsten Borides. *Phys. Rev. Lett.* **110**, 136403 (2013).

[45] Duan, D. F. *et al.* Pressure-induced decomposition of solid hydrogen sulfide. *Phys. Rev. B* **91**, 180502 (2015).

[46] Duan, D. F. *et al.* Pressure-induced metallization of dense (H$_2$S)$_2$H$_2$ with high-Tc superconductivity. *Sci. Rep.* **4**, 6968 (2014).

[47] Wang, Y.C., Lv, J., Zhu, L. & Ma, Y.M. CALYPSO: A method for crystal structure prediction. *Comput. Phys. Commun.* **183**, 2063 (2012).

[48] Wang, Y.C., Lv, J., Zhu, L. & Ma, Y.M. Crystal structure prediction via particle-swarm optimization. *Phys. Rev. B* **82**, 094116 (2010).

[49] Pickard, C. J. & Needs, R J., High-pressure phases of silane. *Phys. Rev. Lett.* **97**, 045504 (2006).

[50] Pickard, C. J. & Needs, R. J., Ab initio random structure searching. *J. Phys.: Condens. Matter* **23**, 053201 (2011).

[51] Perdew, J. P., Burke, K. & Ernzerhof, M. Generalized gradient approximation made simple. *Phys. Rev. Lett.* **77**, 3865 (1996).

[52] Kresse, G. & Furthmuller, J. Efficient iterative schemes for ab initio total-energy calculations using a plane-wave basis set. *Phys. Rev. B* **54**, 11169 (1996).

[53] Blochl, P. E., Projector Augmented-Wave Method. *Phys. Rev. B* **50**, 17953 (1994).

[54] Togo, A., Oba, F. & Tanaka, I., First-principles calculations of the ferroelastic transition between rutile-type and $CaCl_2$-type $SiO_2$ at high pressures. *Phys. Rev. B* **78**, 134106 (2008).

[55] Blaha, P., Schwarz, K., Madsen, G. K. H., Kvasnicka D. & Luitz, J. *WIEN2k*, An Augmented Plane Wave Plus Local Orbitals Program for Calculating Crystal Properties (TU Vienna, Vienna, Austria, 2001).

[56] Mao, H. K., Xu, J. & Bell, P. M. Calibration of the ruby pressure gauge to 800 kbar under quasi-hydrostatic conditions. *J. Geophys. Res.* **91**, 4673 (1986).

[57] Akahama, Y. & Kawamura, H. High-pressure raman spectroscopy of diamond anvils to 250 GPa: Method for pressure determination in the multimegabar pressure range. *J. Appl. Phys.* **96**, 3748 (2004).

**Figure Legends.**

**FIG.1. Energy stability and crystal structures.** (a) Enthalpy curves (relative to the ambient pressure distorted-3R phase) of the interested structures as a function of pressure. The inset shows the calculated enthalpy curves of newly found distorted-1T (relative to the distorted-3R) using different exchange correlation functionals. Pressure-induced symmetrization of the distorted-3R (b) and distorted-1T structures (c).

**FIG.2. The simulated XRD patterns for ReS$_2$ phases compared with experiments.** Blue line represents the distorted-3R (the known ambient structure) and red line presents the distorted-1T (new structure predicted in this work) ReS$_2$ at 11 and 50 GPa compared with the experimental data (black line) from Ref. 36, where we use the same x-ray wavelength (λ=0.4959Å) as experiments. Our predicted distorted-3R agrees very well with the experiments.

**FIG.3. The electronic band structures and density of states (DOS).** (a) The total electronic DOS of various structures at 15 GPa, which clearly shows the distortions in distorted-3R and distorted-1T structures introduce band gaps, which can be attributed to the Peierls mechanism. (b) The total and partial DOS for the distorted-1T ReS$_2$ at 15 GPa. The Fermi level is set to zero. (c) The first Brillion zone and (d) the calculated band structures of the distorted-1T ReS$_2$ at several pressures. The distorted-1T undergoes a metallization transition with a band overlap mechanism upon compression.

**FIG.4. The crystal structure, band and electron-phonon coupling of the $I4_1/amd$ ReS$_2$.** (a) The crystal structure of $I4_1/amd$ at 100 GPa (b) The first Brillion zone and (c) the calculated band structure of $I4_1/amd$ ReS$_2$ at 100 GPa. (d) The calculated phonon dispersions of $I4_1/amd$ ReS$_2$ at 100 GPa. The size of the dots represents the phonon linewidth of each mode. Partial phonon DOS, Eliashberg EPC spectral function a$^2$F($\omega$) and the electron-phonon integral λ($\omega$) are also shown on the right panel.

**FIG.5. The temperature-dependence of the in-plane electric resistance of ReS$_2$ at different pressures.** In-plane resistance as a function of temperature in ReS$_2$ at various pressures: (a) a semi-logarithmic scale below 32.0 GPa, (b) around 70 GPa, showing pressure-induced metallic conductivity, and (c) above 91 GPa, suggesting pressure-induced superconducting characteristic.

**FIG 6.** (a) Isothermal resistance at various pressures on a semi-logarithmic scale at 5 K, 100 K, and 200 K, respectively. (b) Magnetic field dependence of the resistance drop in ReS$_2$ at 102.0 GPa. The inset shows the temperature dependence of the upper critical field $\mu_0 H_{c2}$ at 102.0 GPa. Here, $T_c$ at different magnetic field is determined by the crossing of two lines. The solid line represents the fitting curves based on the GL formula.

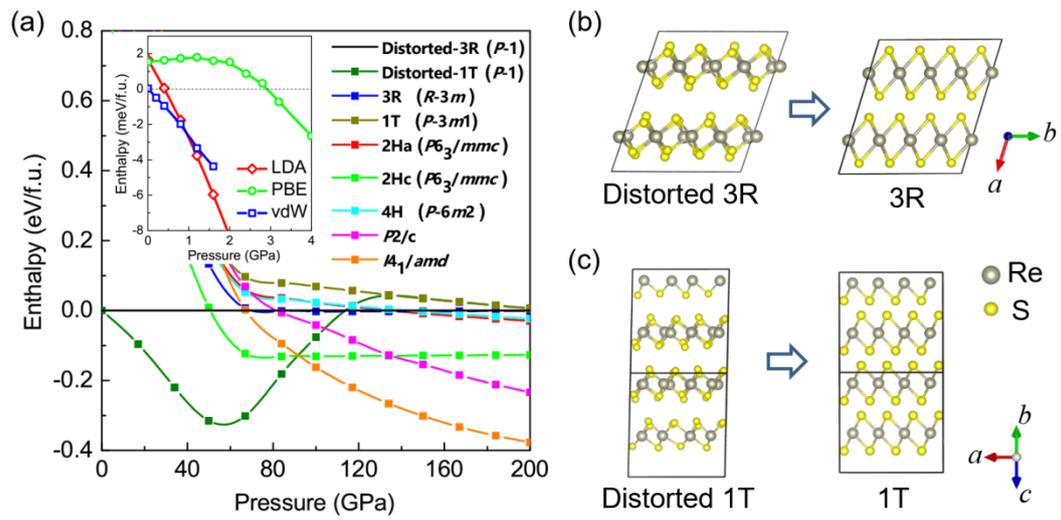

Fig. 1 Zhou *et al*.

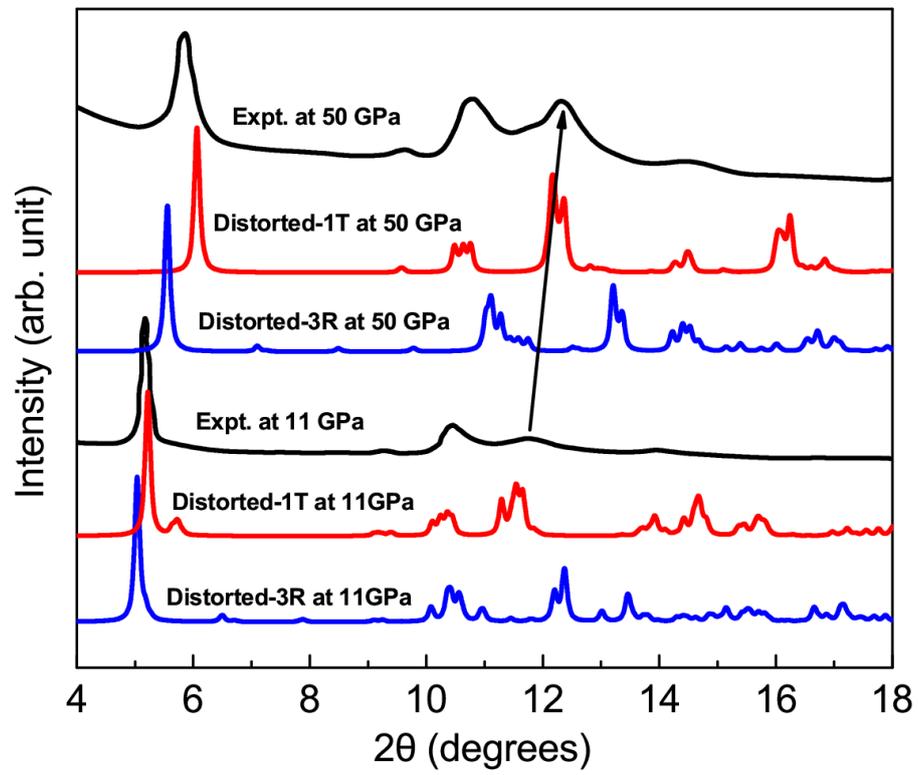

Fig. 2 Zhou *et al*.

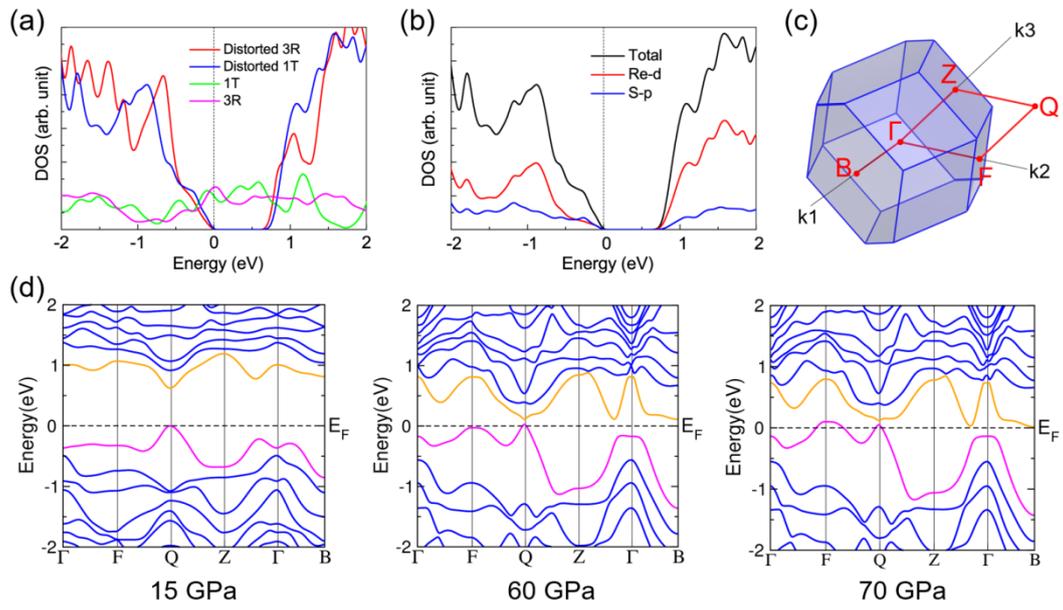

Fig. 3 Zhou *et al*.

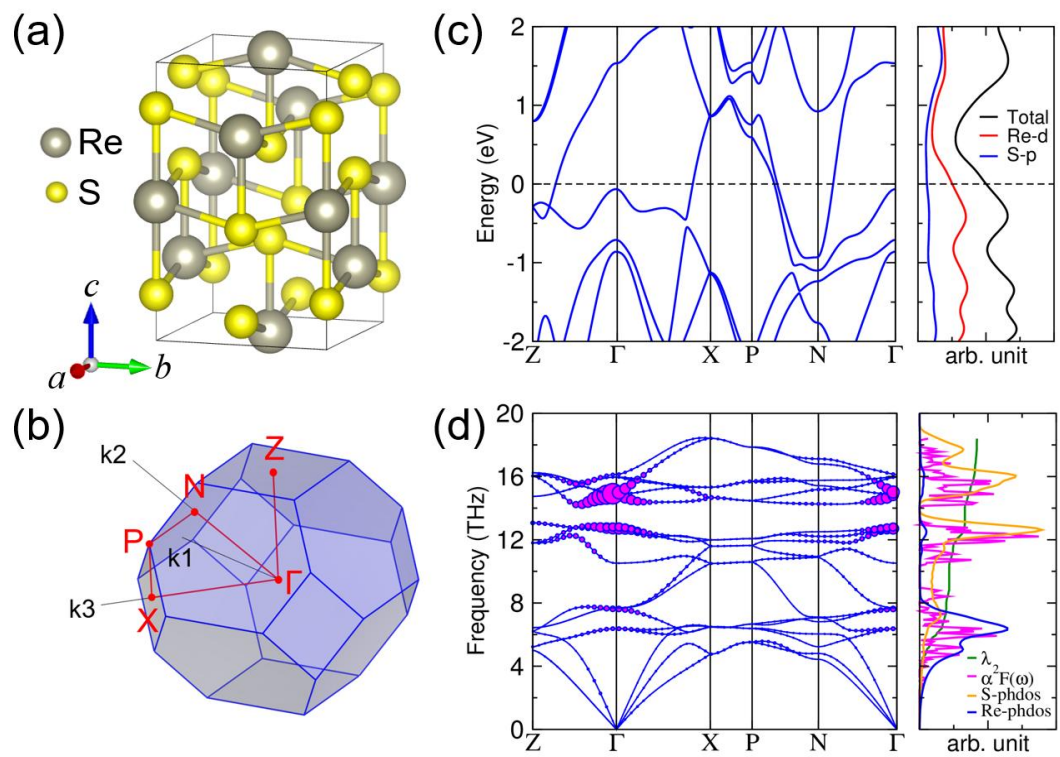

Fig. 4 Zhou *et al*.

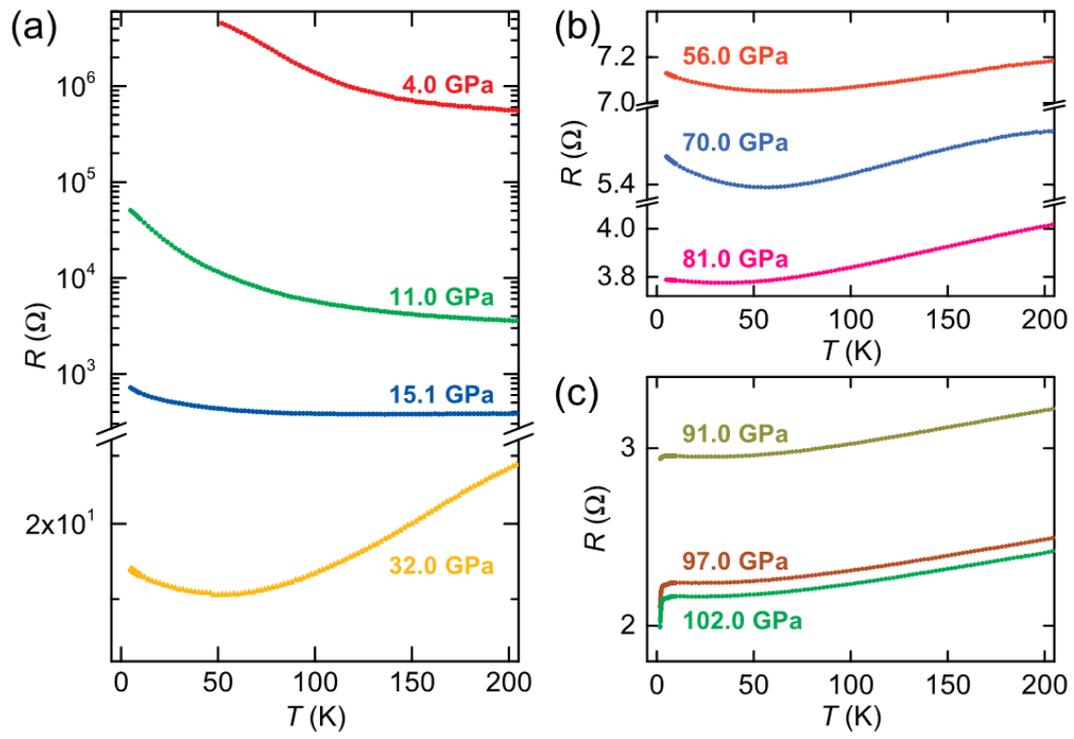

Fig. 5 Zhou *et al*.

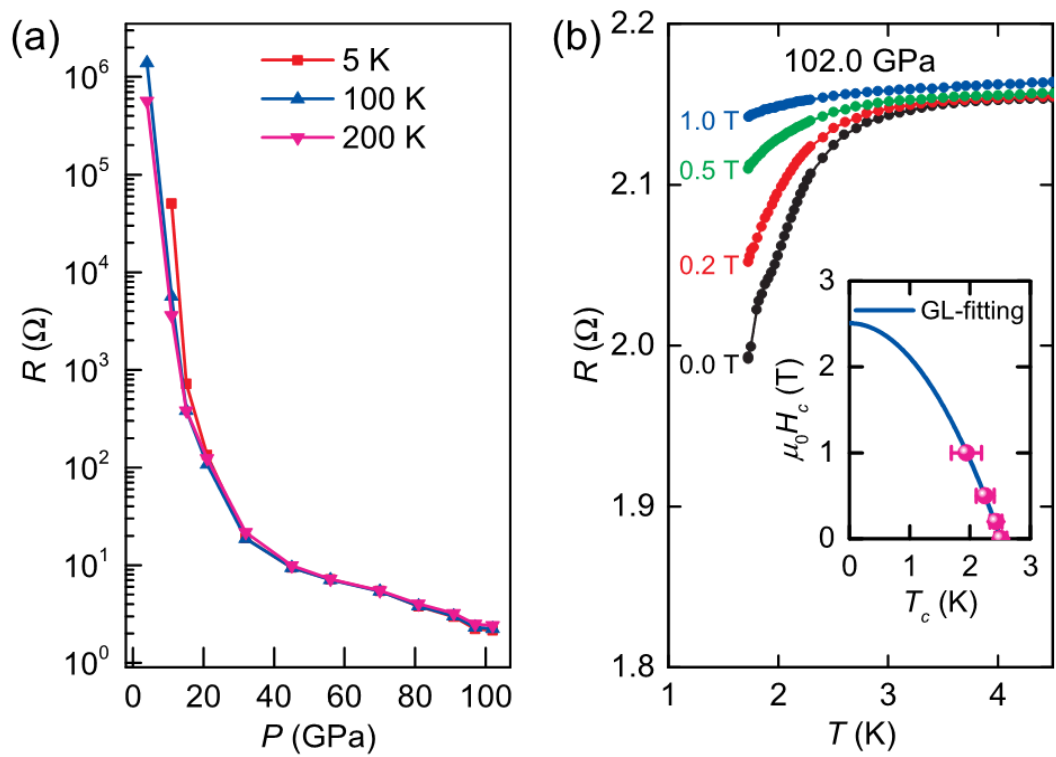

Fig. 6 Zhou *et al*.